\documentclass{amsart}

\usepackage[utf8]{inputenc}
\usepackage[english]{babel}
\usepackage[T1]{fontenc}
\usepackage{amsfonts, amsmath, amssymb, bbm}
\usepackage{amsthm}
\usepackage{xcolor} 

\usepackage{calc}
\usepackage{hhline}
\usepackage{multirow}
\usepackage{setspace}
\usepackage{textcomp, tabulary, stmaryrd}

\usepackage{graphicx}

\newtheorem{theorem}{Theorem}[section]

\theoremstyle{definition}
\newtheorem{ass}[theorem]{Assumption}
\newtheorem{definition}[theorem]{Definition}
\newtheorem{example}[theorem]{Example}
\newtheorem{lem}[theorem]{Lemma}
\newtheorem{paradigm}[theorem]{Paradigm}
\newtheorem{prop}[theorem]{Proposition}

\newtheorem{remark}[theorem]{Remark}

\newenvironment{bew}{\textit{Proof}.}{\hfill$\Box$}

\newcommand{\bes}{\ensuremath{\mathrm{Bes^3}}}
\newcommand{\bib}[4]{#1 (#2). #3. \textit{#4}.}

\newcommand{\eps}{\ensuremath{\varepsilon}}

\newcommand{\F}{\mathbb{F}}

\renewcommand{\i}{\ensuremath{\mathbbm{1}}}
\newcommand{\integ}[4]{\ensuremath{\int_{#1}^{#2}#3\,d#4}}
\newcommand{\leli}{\ensuremath{\textrm{--}}}

\newcommand{\M}[1]{\mathcal{#1}}
\newcommand{\N}{\mathbb{N}}

\newcommand{\Q}{\mathbb{Q}}
\newcommand{\R}{\mathbb{R}}
\newcommand{\set}[2]{\ensuremath{\big\{#1\,\big|\,#2\big\}}}

\renewcommand{\tilde}[1]{\widetilde{#1}}

\DeclareFontFamily{U}{mathx}{\hyphenchar\font45}
\DeclareFontShape{U}{mathx}{m}{n}{
      <5> <6> <7> <8> <9> <10>
      <10.95> <12> <14.4> <17.28> <20.74> <24.88>
      mathx10
      }{}
\DeclareSymbolFont{mathx}{U}{mathx}{m}{n}
\DeclareFontSubstitution{U}{mathx}{m}{n}
\DeclareMathAccent{\widecheck}{0}{mathx}{"71}
\DeclareMathAccent{\wideparen}{0}{mathx}{"75}

\renewcommand{\check}[1]{\ensuremath{\widecheck{#1}}}
\renewcommand{\i}{\ensuremath{\mathbbm{1}}}

\renewcommand{\tilde}{\widetilde}

\allowdisplaybreaks

\begin{document}

\hyphenpenalty=10000

\title{A constraint-based notion of illiquidity}


\author{Thomas Krabichler}
\address{}
\curraddr{}
\email{thomas.krabichler@ost.ch}
\thanks{}

\author{Josef Teichmann}
\address{}
\curraddr{}
\email{josef.teichmann@math.ethz.ch}
\thanks{}


\date{}

\begin{abstract}
This article introduces a new mathematical concept of illiquidity that goes hand in hand with credit risk. The concept is not volume- but constraint-based, i.e., certain assets cannot be shorted and are ineligible as num\'eraire. If those assets are still chosen as num\'eraire, we arrive at a two-price economy. We utilise Jarrow \& Turnbull's foreign exchange analogy that interprets defaultable zero-coupon bonds as a conversion of non-defaultable foreign counterparts. In the language of structured derivatives, the impact of credit risk is disabled through quanto-ing. In a similar fashion, we look at bond prices as if perfect liquidity was given. This corresponds to asset pricing with respect to an ineligible num\'eraire and necessitates Föllmer measures.
\end{abstract}

\maketitle


\section{Introduction}

Let us assume that a financial agent sells buckets and promises to fill each of them with one litre of water in three months. If one buys such a claim, then, amongst others, one is exposed to the following somewhat interlinked risks. First and foremost, there is the risk that the supplier cannot honour their obligations at maturity for whatever reason. For instance, there may not be enough water around or the buckets may not prove to be watertight.

Furthermore, let us assume that the only water supply is from a single lake and that the lake is frozen in its entirety due to severe weather conditions. As a consequence, the water cannot be delivered in due time and only after having invested energy in order to liquefy the desired amount. During the lifetime of the contract, there are several reasons why such a bucket may change hands. For example, one is fortunate, and the extra amount of water is redundant. In any case, a bucket holder runs the risk that the recoverable value is adversely affected, and significant discounts have to be accepted when reselling the claim. Even though one is dealing with a bona fide counterparty and the buckets are of high quality, water might not be demanded for in the market at all. What is more, sound advice or rumours may circulate that the water supplier is not trustworthy and not as reliable in delivering the promised amount as initially expected. Therefore, one wants to get rid of the claim as quickly as possible in terms of a fire sale. In both scenarios, the recoverable value of the bucket is typically reduced substantially.

These considerations are symptomatic for every financial contract. Broadly speaking, one suffers a loss because either something is not backed sufficiently or something is not readily available due to a lack of liquidity. Aspects of liquidity are diverse and relate to both the market and the commodity itself. Regarding interest rate modelling, the commodity of interest is usually money. Despite the fact that liquidity is an intuitive concept, which we all know from our day-to-day experience, it appears to be tricky to translate it into a sound mathematical framework. A centrepiece of this article is an attempt to clarify this notion.

\emph{Credit risk} refers to possible financial losses that a holder of a claim may suffer over the lifetime of the contract. Indisputably, there is an intricate connection between credit risk and aspects of \emph{liquidity}. The IMF Working Paper WP/02/232 defines four different but intertwined terms of liquidity; see \cite{imf} for further details. Predominantly relevant for quantitative risk modelling are two of them, namely the \emph{asset liquidity}, describing the immediacy and the transaction cost with which the asset in scope can be converted into legal tender, and the \emph{institutional liquidity}, standing for the issuer's ability to meet their settlement obligations.
The promise of a risky loan can be understood, amongst others, as the lender's compensation for being exposed to \emph{inflation risk}, institutional liquidity risk of the issuer and asset liquidity risk.
\begin{itemize}
\item The inflation risk is caused by the time value of money. The notional's purchasing power might weaken over the loan's lifetime and a lender wants to be compensated for that risk.
\item The institutional liquidity risk, more commonly known as default risk, is the event of a liquidity squeeze at expiry or even a premature insolvency of the debtor party implicating only a partial or zero recovery of the loan's face value.
\item The asset liquidity risk comprises the fact that the lenders forgo their own institutional liquidity over the lifetime of the product. If they face an intermediary liquidity squeeze and a fire sale is their last resort, they might have to accept significant discounts on the fair value of the loan. Noteworthy, the notion of asset liquidity solely relates to the loan itself but not to the liquidity of the issuer's assets.
\end{itemize}

Filipovi\'{c} and Trolle corroborate in \cite{interbank} that, subsequent to the credit crunch in 2007/2008, asset liquidity constituted a significant fraction of the risk compensation in the money market. From the phenomenological viewpoint, asset illiquidity involves two prices for a certain good. One price is the \emph{fundamental value}, which is the intrinsic economic value or the minimal cost to replicate this product as if there was perfect liquidity. The other price is a \emph{market value} which is derived from transactions. One typically has to accept a certain discount when converting an illiquid good into cash. The resulting difference between the fundamental value and the market value cannot be exploitet. Below, we translate this observation into a rigorous mathematical statement; illiquidity causes the alleged arbitrage opportunity to be inadmissible. We analyse interest rate modelling in the presence of illiquidity by exploiting a neat foreign exchange (FX) modelling framework.

The \emph{FX-analogy} was originally introduced in a working paper by Jarrow \& Turnbull. A comprehensive exposition can be found in \cite{krabi}, in particular also in a two-filtration setting. This framework enables a joint modelling framework for institutional liquidity and asset liquidity that goes hand in hand with credit risk. Here we consider the Jarrow \& Turnbull setting with only one filtration. However, the exchange rate together with the foreign bank account is only a local martingale in general (if discounted by the domestic bank account). This situation can be categorised in four different ways depending on the properties of the discounted foreign bank account. It can constitute different sorts of liquidity crises in the market, since the foreign bank account can possibly not be taken as a num\'eraire without changing price structures. We therefore arrive at a two-price economy depending on whether we discount in domestic or foreign terms. We do introduce arbitrage opportunities in the market if one is allowed to short the foreign bank account.

Throughout the article, we use the following notion of liquidity.

\begin{definition}[Liquidity and Liquidity Risk]
\emph{Liquidity} is an entity's ability to incur debts immediately. A possible lack of liquidity, affecting issued loans and the entity's solvency likewise, is referred to as \emph{liquidity risk}.
\end{definition}

As illustrated by \cite{gmz}, asset liquidity in modern financial markets is a key but elusive concept. The above notion of liquidity was proposed by the authors in \cite{krabi} and pursues the idea of \cite{kst2013}. In their introductory section of \cite{lehalle}, Lehalle and Laruelle define illiquidity for a specific demanded quantity as round trip cost. This concept can be linked to the proposed theory below by extending the idea to roll-over strategies; see Section~5.2 in \cite{krabi}. Ruf and others generalised in \cite{cfr} and \cite{pr} the change of num\'eraire technique to dominating Föllmer measures in order model hyperinflation in multi-currency settings. This was further substantiated in \cite{defnum} in the context of defaultable num\'eraires. Chau and Tankov study the same setting in \cite{chautank} for optimal arbitrage. In a different direction goes the liquidity risk approach by \c{C}etin, Jarrow and Protter; see \cite{cjp}. They model prices for a single default-free asset, e.g., for a zero-coupon bond with a fixed maturity, both time- and order-size-dependent. This leads to the concept of a \emph{supply curve} which characterises the composition of the order book at a given time instance. Similarly, Madan studies two-price economies in \cite{madan} in order to account for other risks such as liquidity.

The article is structured as follows. In Section~\ref{sec:jt}, we recall the famous FX-analogy by Jarrow \& Turnbull. In the Sections~\ref{sec:kst} and \ref{sec:nutshell}, we present the key ideas without going into any technical details. The necessary technical toolkit is derived in the Sections~\ref{sec:föllmer} and \ref{sec:basis}. The centrepiece of this article is Section~\ref{sec:illprem}, where we present four distinct market scenarios for aspects of liquidity.

\section{The Jarrow \& Turnbull Setting}\label{sec:jt}

Let $[0,\infty)$ be the considered timeline. We denote by $\big(P(t,T)\big)_{0\leq t\leq T}$ the c\`adl\`ag price process of a non-defaultable zero-coupon bond with maturity $T\geq 0$ and payoff $P(T,T)=1$. Furthermore, we denote by $\big(\tilde{P}(t,T)\big)_{0\leq t\leq T}$ the c\`adl\`ag price process of a defaultable zero-coupon bond with the same maturity and a random payoff $0<\tilde{P}(T,T)\leq 1$. We assume that $P(T,T)$ and $\tilde{P}(T,T)$ are written in the same currency. The distribution of the final recovery $\tilde{P}(T,T)$ is strongly linked to the riskiness of the issuer's business model. It needs to be noted that we do not allow the final payoff to become zero and we make this assumption for any maturity $T\geq 0$. We shall see straightaway why this assumption is of crucial importance for our modelling approach. Consequently, we may introduce another term structure $\big\{Q(t,T)\big\}_{0\leq t\leq T<\infty}$ via $Q(t,T):=\frac{\tilde{P}(t,T)}{\tilde{P}(t,t)}$. Note that we have $Q(T,T)=1$ and, hence, that this synthetic series is default-free. By setting $S_t:=\tilde{P}(t,t)$, we get $\tilde{P}(t,T)=S_tQ(t,T)$. Although this rewriting is very elementary, it opens an extremely nice modelling opportunity for defaultable zero-coupon bonds. We recognise that credit risk can be analysed in an FX-like setting.

\begin{paradigm}[Jarrow \& Turnbull 1991]\label{par:jt91}
The series $P(t,T)$ and $Q(t,T)$ are considered as non-defaultable zero-coupon bonds in different currencies. $\tilde{P}(t,T)$ may be interpreted as conversion of foreign default-free counterparts. $S_t=\tilde{P}(t,t)$ is referred to as \emph{recovery rate} or \emph{spot FX rate}.
\end{paradigm}

In order to utilise the FX-analogy from a mathematical perspective, we fix the following setup.

\begin{ass}[The General FX-like Setting]\label{ass:gen}
Let $(\Omega,\M{F},\F,\Q)$ with $\F=(\M{F}_t)_{t\geq 0}$ be a filtered probability space satisfying the usual conditions. We consider $\Q$ as risk-neutral pricing measure. By $B=(B_t)_{t\geq 0}$ we describe the accumulation of the domestic risk-free bank account with initial value of one monetary unit and by $\check{B}=(\check{B}_t)_{t\geq 0}$ its foreign counterpart. Furthermore, let $\big\{P(t,T)\big\}_{0\leq t\leq T<\infty}$, $\big\{\tilde{P}(t,T)\big\}_{0\leq t\leq T<\infty}$ and $\big\{Q(t,T)\big\}_{0\leq t\leq T<\infty}$ be three $\F$-adapted families capturing the stochastic evolution of the term structure of zero-coupon bond prices. More precisely, $\omega\longmapsto P\big(t,T\big)(\omega)$, $\omega\longmapsto \tilde{P}\big(t,T\big)(\omega)$ and $\omega\longmapsto Q\big(t,T\big)(\omega)$ are supposed to be positive and $\M{F}_t$-measurable a.s.\ for all $0\leq t\leq T<\infty$. Additionally, the mappings $t\longmapsto P(t,T)$, $t\longmapsto\tilde{P}(t,T)$ and $t\longmapsto Q(t,T)$ are supposed to have c\`adl\`ag paths a.s.\ for all $0\leq T<\infty$. The corresponding payoffs satisfy
\begin{equation}\label{def:payoffp} P(T,T)=1,\qquad 0<S_T:=\tilde{P}(T,T)\leq 1\end{equation}
in the domestic currency, and the relation
\begin{equation}\label{def:payoffq}Q(t,T)=\frac{\tilde{P}(t,T)}{S_t}\end{equation}
in some synthetic foreign currency. We assume the three properties in \eqref{def:payoffp} and \eqref{def:payoffq} to hold a.s.\ for all $0\leq t\leq T<\infty$. Finally, we assume absence of arbitrage in the sense that the discounted price processes
\[\frac{P(t,T)}{B_t},\qquad\frac{S_tQ(t,T)}{B_t}=\frac{\tilde{P}(t,T)}{B_t},\qquad\frac{S_t\check{B}_t}{B_t}\]
for $0\leq t\leq T$ are $\Q$-local martingales for each $T\geq 0$.
\end{ass}

The FX-like setting provides a powerful machinery to study credit and liquidity risk; see \cite{krabi} or \cite{krabiteichm} for further details and examples.

\section{A First Step Towards Modelling Illiquidity}\label{sec:kst}

In many articles on mathematical finance, it is assumed inherently that entering an \emph{arbitrarily large} short position in the num\'eraire is admissible. After the existence of a bank account, this is yet another very strong assumption. Despite being controversial, this finding may serve as the basic idea to model consequences of asset liquidity constraints. All we have to do is to reverse the rationale. \emph{A financial asset subject to asset liquidity constraints cannot be shorted arbitrarily. Hence, it cannot serve as num\'eraire either.} See also Lemma~4.12 and Proposition~4.13 in \cite{krabi}. We are going to illustrate the basic idea by the following example, which is lent from the introductory section of \cite{kst2013}. This article links the concept of \emph{illiquidity} to \emph{ineligible num\'eraires} and \emph{bubbles}; see Theorem~2.8 in \cite{kst2013}. We will formalise this idea in the subsequent Section~\ref{sec:illprem}. Meanwhile, we prepare the necessary technical building blocks. 

\begin{example}[Klein, Schmidt \& Teichmann 2013]\label{bsp:kst}
\textup{
We consider the general FX-like setting of Assumption~\ref{ass:gen} and fix some maturity $T>0$. We assume that $\Q$ is a true martingale measure for $\big(\tilde{P}(t,T)\big)_{0\leq t\leq T}$. Additionally, we assume full initial recovery $S_0=1$ and that $\M{F}_0$ contains only trivial information. We define the $\Q$-local martingale $Z=(Z_t)_{0\leq t\leq T}$ as $Z_t:=\frac{S_t\check{B}_t}{S_0B_t}$. If $Z$ is a true $\Q$-martingale, then $Z$ represents a density process for some equivalent measure $\check{\Q}\approx\Q$, and the classical change of num\'eraire technique says $Q(0,T)=E_{\check{\Q}}\big[\frac{1}{\check{B}_T}\big]$. If $Z$ is a strict $\Q$-local martingale, hence also a strict $\Q$-supermartingale with $E_\Q\big[Z_T\big]<1$, then we may still define a locally equivalent measure $\check{\Q}\big|_{\M{F}_T}\approx\Q\big|_{\M{F}_T}$ via
\[\frac{d\check{\Q}}{d\Q}\bigg|_{\M{F}_T}:=\frac{Z_T}{E_\Q\big[Z_T\big]}.\]
Consistently, we get
\[\check{Q}(0,T):=E_{\check{\Q}}\bigg[\frac{1}{\check{B}_T}\bigg]=\frac{1}{E_\Q\big[Z_T\big]}E_\Q\bigg[\frac{Z_T}{\check{B}_T}\bigg]=\frac{1}{E_\Q\big[Z_T\big]}Q(0,T)>Q(0,T),\]
i.e., we end up in a two-price economy. The price of a defaultable zero-coupon bond $S_0\check{Q}(0,T)$ with respect to the num\'eraire $\check{B}$ and the pricing measure $\check{\Q}$ is higher than $\tilde{P}(0,T)=S_0Q(0,T)$ in the initial market. Nonetheless, we proclaim that the price difference cannot be exploited due to asset liquidity constraints. In this context, $Z$ is not considered as being an eligible num\'eraire. While being a strict local martingale, $Z$ exceeds any barrier with positive probability. In order to emulate a replicating strategy and build the synthetic asset consisting of a long position in $Q(0,T)$ and, more crucially, a short position in $\check{Q}(0,T)$, an arbitrarily negatively valued short position in either finite or infinitesimal roll-overs of defaultable zero-coupon bonds would be required; and partial recovery must not hold over the replication period either. As indicated by Definition~6.1 in \cite{kst2013}, the replication involves the reciprocal value of $\tilde{P}(t,t+dt)$. Analogously, the strict local martingale property of $Z$ features the market phenomenon that no market participant is willing to lend capital based on entering a \emph{repurchase agreement} (repo) with this synthetic asset. The bubble may burst any time and devalue the collateral strongly. This is synonymous with a contingent or qualified interest in holding defaultable zero-coupon bonds; see also Example~\ref{bsp:explicit} below for further clarification.}\hfill$\Box$
\end{example}

Note that the above argument only works for the time instance $t=0$ and is somehow maturity-dependent. In contrast to Example~\ref{bsp:kst}, we do not charge the payoff $1$ with the defect of $Z_T$. Instead, we put mass into a hidden default, which can only be seen under the new measure $\check{\Q}$. Correspondingly, the enhanced setting founds on the existence of a foreign bank account together with an associated dominating pricing measure.

The above concept of liquidity raises several discussion points. In mathematical terms, it can hardly be analysed on a stand-alone basis. Once credit risk has been disabled and $S_t\equiv 1$ prevails, one will end up in a pathological model, in which $\tilde{P}(t,T)=P(t,T)$ holds for all $0\leq t\leq T<\infty$. Seen from time $t$ and with respect to a selected num\'eraire, there must be a unique value for the payoff of one monetary unit at time $T$. Unless shorting either of the zero-coupon bonds is not admissible, any discrepancy between $P(t,T)$ and $\tilde{P}(t,T)$ could be exploited as a free lunch with a simple buy-and-hold strategy. To this extent, asset liquidity can hardly be isolated. In a financial model, one cannot contrast two financial assets equipped with the identical payoff but two different levels of liquidity. Contrasting is the natural approach for studying credit risk; see Section~\ref{sec:jt}. It needs to be noted that this statement is only affecting financial modelling. In the real world, there might be coequal bonds within the same discrete rating class but deviating yields. From the modelling viewpoint, the impact of liquidity can only be uncovered by a change of num\'eraire together with an associated change of measure. The \emph{marketable price} $\tilde{P}(t,T)$ is with respect to some objective num\'eraire. In our case, this is normally the bank account $B=(B_t)_{t\geq 0}$. The \emph{fundamental value} or \emph{intrinsic economic value} $S_t\check{Q}(t,T)$ is with respect to some synthetic num\'eraire $\check{B}=(\check{B}_t)_{t\geq0}$ capturing the assumed accumulation of returns, if roll-overs of the considered financial asset were possible. The resulting price difference is the premium solely attributed to asset liquidity; see also Remark~\ref{rmk:mathilliq} below.

\section{The Long Story Short}\label{sec:nutshell}

Conceptually, in the centre of the enhanced FX-like setting lies a process $Z$ that is either exogenously given or the result of a roll-over strategy in defaultable zero-coupon bonds. Due to asset liquidity constraints, this portfolio value process is in a bubble state under a risk-neutral measure $\Q$. The bubble state is embodied by the strict local martingale property of $Z$, whose characteristic sample paths show a hump with a far end being below its long-term mean. Consequently, no counterparty is willing to accept it as collateral in the context of lending money. The bubble could burst any time and devalue the collateral. If one believes in the bubble state, one is tempted to short the asset and take advantage of the anticipated price collapse. However, in mathematical terms, shorting $Z$ is not admissible under $\Q$, since $-Z$ is not a $\Q$-local martingale. The presence of the bubble goes hand in hand with assigning zero probability to a \emph{liquidity event}.

A dominating Föllmer measure $\check{\Q}$ for $Z$, as introduced in the next section, enables to generalise the change of num\'eraire technique. Under $\check{\Q}$, the discounted value of the num\'eraire evolves flat on the level one. Thus, the bubble is not visible under $\check{\Q}$ and $Z_0=1$ seems to be priced correctly. Allowing $Z$ as num\'eraire implies a new pricing regime. The change in credit lines gives reason for new superreplication prices. Generally, as pinpointed by the model of the $4^\textrm{th}$ kind below, this happens without order relation, i.e., prices can either rise or fall compared to the initial setting. $Z$ is the natural num\'eraire in the economic pricing of the defaultable zero-coupon bonds and determines the so-called fundamental value. It treats them as if perfect asset liquidity was given. Since there is no order relation in the prices, the illiquidity premium can attain both signs. Though, illiquidity appears more natural than hyperliquidity. The latter is a pathological phenomenon of imperfect markets. An entirely positive illiquidity premium is easily achievable by means of the model of the $2^\textrm{nd}$ kind. The scenario in which no one is willing to hold the defaultable zero-coupon bonds under any circumstances whatsoever translates into the promised yield exceeding any rational level. The abrupt devaluation leads to a hyperinflation in the foreign market and $Z$ explodes. As seen under $\Q$, the liquidity event is featured in terms of a hidden (i.e., improbable) default of foreign zero-coupon bonds occurring at the stopping time $\tau$. It is as if one took the equity of the considered issuer as num\'eraire. The singularity occurring at time $\tau$ refers to equity on the edge of becoming negative.

Under mild technical assumptions, the financial market under $\check{\Q}$ is arbitrage-free; see Lemma~\ref{lem:naqcheckb}. In the original market with respect to $\Q$, the difference between market prices and fundamental prices cannot be exploited due to admissibility constraints. If these were ignored, one could materialise the discrepancy. An optimal arbitrage profit would lurk in the self-financing, yet inadmissible, replication of $\i_{\{\tau>T\}}$ with $B$ and $Z$.

We utilise the quanto-ing technique from Jarrow \& Turnbull in order to analyse institutional liquidity risk; see Paradigm~\ref{par:jt91}. To this end, we somehow disable default risk through introducing the foreign term structure $T\longmapsto Q(t,T)$. For asset liquidity risk, we go the other way round. By changing the num\'eraire from $(\Q,B)$ to $(\check{\Q},Z)$, we discover a previously unlikely default feature $Q(T,T)=\i_{\{\tau> T\}}$. From a risk management perspective, new scenarios are added in order for absence of arbitrage to prevail. Essentially, institutional liquidity risk and asset liquidity risk are modelled jointly in a similar way and interact with each other. This can be perceived as a duality of credit and liquidity risk.

\section{Föllmer Measures}\label{sec:föllmer}

The following probabilistic exposition is inspired by \cite{cfr}, \cite{imkellerperk}, \cite{larsson} and \cite{pr}. Under suitable conditions, supermartingales may be seen as generalised density processes. The importance of Föllmer measures in the context of mathematical finance can be recognised by Theorem~4.14 in \cite{imkellerperk}. Let $(\Omega,\M{F},\F,\Q)$ with $\F=(\M{F}_t)_{t\geq 0}$ be a filtered probability space. 

\begin{definition}[Standard System]
The filtration $\F$ is called a \emph{standard system}, if $(\Omega,\M{F}_t)$ is isomorphic to some separable complete metric space with its Borel $\sigma$-algebra for each $t\geq 0$, and if it holds $\bigcap_{n\in\N}A_n\neq\varnothing$ for all non-decreasing sequences $(t_n)_{n\in\N}$ and for all non-increasing sequences of atoms $(A_n)_{n\in\N}$ with $A_n\in\M{F}_{t_n}$ for each $n\in\N$.
\end{definition}

We assume that $\F$ is the right-continuous modification of a standard system. Since we are going to work with dominating local martingale measures, we do not augment $\F$ with the $\Q$-nullsets. See Lemma~\ref{lem:Fnotcomplete} below, or \cite{imkellerperk}, \cite{larsson} and \cite{pr} for further details.

Let $Z=(Z_t)_{t\geq 0}$ be a non-negative $\Q$-local martingale with $Z_0=1$ and c\`adl\`ag paths. We define the stopping times
\begin{equation}\label{eq:explosion}\tau_n:=n\wedge\inf\set{t\geq 0}{Z_t> n},\qquad\tau:=\lim_{n\to\infty}\tau_n.\end{equation}
Each $\tau_n$ for $n\in\N$ is the \emph{capped hitting time} of an open set and, hence, an $\F$-stopping time according to Lemma~6.6 (iii) in \cite{kallenberg}. $\tau$ also is an $\F$-stopping time since
\[\{\tau\leq t\}=\bigcap_{\eps\in\Q\cap(0,\infty)}\bigcup_{m\in\N}\bigcap_{n=m}^\infty\big\{\tau_n\leq t+\eps\big\}\]
and $\F$ is assumed to be right-continuous. If $(\sigma_n)_{n\in\N}$ denotes a localising sequence of $Z$, then it holds by Fatou's Lemma for all $0\leq u\leq t\leq\infty$
\[E_\Q\big[Z_t\big|\M{F}_u\big]=E_\Q\Big[\liminf_{n\to\infty}Z_{t\wedge\sigma_n}\Big|\M{F}_u\Big]\leq\liminf_{n\to\infty}E_\Q\big[Z_{t\wedge\sigma_n}\big|\M{F}_u\big]=Z_u.\]
Therefore, $Z$ is a $\Q$-supermartingale. Setting $u=0$ and taking expectations on both sides of the above argument yields $E_\Q\big[Z_t\big]\leq 1$ for all $t\geq 0$. This together with the c\`adl\`ag property of $Z$ guarantees $\Q\big[\tau<\infty\big]=0$, i.e., $Z$ does not explode in finite time under $\Q$. According to Section~2 in \cite{larsson}, there exists a unique probability measure $\check{\Q}$, the so-called \emph{Föllmer measure} of  $Z$, on the sub-$\sigma$-field
\[\M{F}_{\tau\leli}:=\sigma\Big(\set{A\cap\{\tau>t\}}{A\in\M{F}_t\textrm{ for some }t\geq 0}\Big),\]
such that it holds
\begin{equation}\label{eq:föllmer1}\check{\Q}\big[A\cap\{\tau>t\}\big]=E_\Q\big[Z_t\i_A\big]\end{equation}
for all $A\in\M{F}_t$ and all $t\geq 0$. Consistently, this may be extended to $E_{\check{\Q}}\big[H_t\i_{\{\tau>t\}}\big]=E_\Q\big[Z_tH_t\big]$ for all $\check{\Q}$-integrable $\M{F}_t$-measurable random variables $H_t$. Particularly, it holds $\check{\Q}\big[\tau=\infty\big]=\lim_{k\to\infty}\check{\Q}\big[\tau>k\big]=\lim_{k\to\infty}E_\Q\big[Z_k\big]$. If we set
\begin{equation}\label{eq:invZ}\check{Z}_t:=\begin{cases}\frac{1}{Z_t}\i_{\{\tau>t\}}&\textrm{, on }\{Z_t>0\},\\0&\textrm{, otherwise,}\end{cases}\end{equation}
then we will get for $A\in\M{F}_t$ and $H_t=\check{Z}_t\i_A$ the inverse transformation formula
\begin{equation}\label{eq:föllmer2}\Q\big[A\cap\{Z_t>0\}\big]=\Q\big[A\cap\{Z_t>0\}\cap\{\tau>t\}\big]=E_{\check{\Q}}\big[\check{Z}_t\i_A\big].\end{equation}
Analogously, we get $E_\Q\big[H_t\i_{\{Z_t>0\}}\big]=E_{\check{\Q}}\big[\check{Z}_tH_t\big]$ for all $\Q$-integrable $\M{F}_t$-measurable random variables $H_t$. In the sequel, the processes $Z=(Z_t)_{t\geq 0}$ and $\check{Z}=(\check{Z}_t)_{t\geq 0}$ are referred to as \emph{generalised density processes}.

Even though Equation~\eqref{eq:föllmer1} characterises $\check{\Q}$ only on $\M{F}_{\tau\leli}$, we want $\check{\Q}$ to be defined on the whole $\sigma$-field $\M{F}$. Consistent with Definition~2.1 and Proposition~2.3 in \cite{pr}, where Föllmer measures are considered from a formal perspective, we make the following definition.

\begin{definition}[Föllmer Pair]\label{def:föllmerpair}
Let $(\Omega,\M{F},\F,\Q)$ with $\F=(\M{F}_t)_{t\geq 0}$ be a filtered probability space, where $\F$ is the right-continuous modification of a standard system, and $Z=(Z_t)_{t\geq 0}$ be a non-negative $\Q$-supermartingale with c\`adl\`ag paths and $Z_0=1$. Furthermore, let $\check{\Q}$ be another probability measure on $\M{F}$ and $\tau$ be a stopping time. Then, $(\check{\Q},\tau)$ is called a \emph{Föllmer pair} for $Z$, if $\Q\big[\tau=\infty\big]=1$ and Equation~\eqref{eq:föllmer1} holds for all $A\in\M{F}_t$ and all $t\geq 0$.
\end{definition}

Theorem~3.1 in \cite{pr} provides an existence and (non-)uniqueness result for Föllmer measures on state spaces; see Definition~C.3 in \cite{pr} for a reference. Beyond the time instance $\tau$, $\check{\Q}$ can be extended arbitrarily without breaking Equation~\eqref{eq:föllmer1}; see also item~(iii) in the Appendix~B of \cite{cfr}.

Given two probability measures $\Q$ and $\check{\Q}$, where $\check{\Q}$ is a Föllmer measure with respect to $Z$, then $\tau$ is uniquely determined up to a $\check{\Q}$-nullset and $Z$ is uniquely determined up to a $\Q$-evanescent set; see Proposition~2 in \cite{yoerp}. If $Z$ is a true $\Q$-martingale, then we also have $\check{\Q}\big[\tau<\infty\big]=0$. In this case, $Z$ becomes the classical \emph{Radon-Nikodym density process} for the locally absolutely continuous measure $\check{\Q}\ll\Q$. More precisely, it holds $\check{\Q}\big|_{\M{F}_t}\ll\Q\big|_{\M{F}_t}$ for all $t\geq 0$. If, in addition, $Z$ is strictly positive $\Q$-a.s., then $\Q$ and $\check{\Q}$ are locally equivalent. If $Z$ is strictly positive $\Q$-a.s.\ but not necessarily a true $\Q$-martingale, then only the local relation $\Q\big|_{\M{F}_t}\ll\check{\Q}\big|_{\M{F}_t}$ is assured for each $t\geq 0$. Generally, there is no order relation in the sense of $\ll$ between $\Q$ and $\check{\Q}$. As highlighted in \cite{pr}, $\Q\big|_{\M{F}_t}$ and $\check{\Q}\big[\cdot\big|\tau\leq t\big]\Big|_{\M{F}_t}$ are even singular, given that $\check{\Q}\big[\tau\leq t\big]>0$. The first one has full mass on the event $\{\tau=\infty\}$, while the other assigns zero mass to it.

\begin{lem}[Generalised Bayes Formula]\label{lem:genbayes}
Consider the setting of Definition~\ref{def:föllmerpair}. Then the following Bayes formula for conditional expectations holds $\check{\Q}$-a.s.
\begin{equation}\label{eq:bayes}\i_{\{Z_t>0\}}E_{\check{\Q}}\big[H_T\i_{\{\tau>T\}}\big|\M{F}_t\big]=\check{Z}_tE_\Q\big[Z_TH_T\big|\M{F}_t\big]\end{equation}
for all $0\leq t\leq T<\infty$ and all $\M{F}_T$-measurable random variables $H_T$, which are both $\Q$- and $\check{\Q}$-integrable.
\end{lem}

\begin{bew}
Let $A\in\M{F}_T$ and $B\in\M{F}_t$. Then, by applying the above transformation formulae \eqref{eq:föllmer1} and \eqref{eq:föllmer2} forth and back, we may write
\begin{align*}
E_{\check{\Q}}\Big[\i_A\i_B\i_{\{\tau>T\}}\i_{\{Z_t>0\}}\Big]&=E_\Q\Big[Z_T\i_A\i_B\i_{\{Z_t>0\}}\Big]\\
&=E_\Q\Big[E_\Q\big[Z_T\i_A\big|\M{F}_t\big]\i_B\i_{\{Z_t>0\}}\Big]\\
&=E_{\check{\Q}}\Big[\check{Z}_tE_\Q\big[Z_T\i_A\big|\M{F}_t\big]\i_B\Big].
\end{align*}
The standard machine from measure theory yields the assertion.
\end{bew}
\\

If $Z$ is strictly positive $\Q$-a.s.\ but not necessarily a true $\Q$-martingale, then Formula~\eqref{eq:bayes} simplifies to
\begin{equation}\label{eq:bayespos} E_{\check{\Q}}\big[H_T\i_{\{\tau>T\}}\big|\M{F}_t\big]=\frac{1}{Z_t}\i_{\{\tau>t\}}E_\Q\big[Z_TH_T\big|\M{F}_t\big],\end{equation}
which holds a.s.\ under $\Q$ and $\check{\Q}$ alike. The choice $t=0$ and $H_T\equiv 1$ in \eqref{eq:bayes} results in the relation $\check{\Q}\big[\tau>T\big]=E_\Q\big[Z_T\big]$. Hence, $T\longmapsto E_\Q\big[Z_T\big]$ describes the distribution of the explosion time under $\check{\Q}$. Equation~\eqref{eq:bayespos} reminds of a popular intensity-based pricing formula when default risk is modelled via filtration enlargement; e.g., see Corollary~7.3.4.2 in \cite{jeanblanc}.

The next lemma describes how to realise the setting of Definition~\ref{def:föllmerpair}. The lemma is based on Proposition~2.5 in \cite{pr} and can be seen as a generalisation of Theorem~1 in \cite{delbaen:bessel}. It is also presented as Theorem~1.1 in \cite{kkn}.

\begin{lem}[Inverse Construction Scheme]\label{lem:icsffm}
For a start, let $(\Omega,\M{F},\F,\check{\Q})$ be a filtered probability space satisfying the usual conditions. Moreover, let $\check{Z}=(\check{Z}_t)_{t\geq 0}$ with $\check{Z}_0=1$ be a non-negative uniformly integrable $(\F,\check{\Q})$-martingale. Define the locally absolutely continuous measure $\Q$ on $\M{F}_\infty:=\bigvee_{t\geq 0}\M{F}_t$ via
\[\frac{d\Q}{d\check{\Q}}\bigg|_{\M{F}_t}:=\check{Z}_t.\]
Converse to \eqref{eq:explosion}, define for each $n\in\N$ the $\F$-stopping times
\[\tau_n:=\inf\set{t\geq 0}{\check{Z}_t<1/n},\qquad\tau:=\lim_{n\to\infty}\tau_n.\]
Then $(\check{\Q},\tau)$ forms a Föllmer pair for the $\Q$-supermartingale $Z:=\check{Z}^{-1}$. In addition, the following two equivalence statements hold:
\begin{itemize}
\item $Z$ is a $\Q$-local martingale if and only if
\begin{equation}\label{eq:notjump}\check{\Q}\big[\tau<\infty,\check{Z}_{\tau\leli}\neq 0\big]=0,\end{equation}
i.e., $\check{Z}$ does not jump to zero $\check{\Q}$-a.s.
\item $Z$ is a true $\Q$-martingale if and only if $\check{Z}$ is strictly positive $\check{\Q}$-a.s.
\end{itemize}
\end{lem}

Noteworthy, the standard bottom-up approach as in \cite{larsson} or \cite{pr}, with a possibly non-unique $\check{\Q}$, is consistent with the proposed top-down construction. According to Lemma~3 in \cite{larsson}, Condition~\eqref{eq:notjump} is satisfied naturally. As is well-known, $\check{Z}$ will stay in zero after $\tau$ $\check{\Q}$-a.s.; e.g., see Proposition~II.3.4 in \cite{revuzyor}.\\

\begin{bew}
It needs to be shown that the following four items hold:
\begin{enumerate}
\item $\Q\big[\tau<\infty\big]=0$.
\item Equation~\eqref{eq:föllmer1} is satisfied for all $A\in\M{F}_t$ and all $t\geq 0$.
\item $Z$ really is a $\Q$-supermartingale.
\item The martingale properties of $Z$ under $\Q$ are equivalent to the stated path properties of $\check{Z}$ under $\check{\Q}$.
\end{enumerate}
We proceed in successive steps: \emph{Proof of 1}. By the right-continuity of $\check{Z}$, we have the upper bound $\check{Z}_{\tau_n}\leq\frac{1}{n}$ on the event where $\tau_n$ is finite. Hence, it holds for all $t\geq 0$ and for all $n\in\N$
\[\Q\big[\tau\leq t\big]\leq\Q\big[\tau_n\leq t\big]=E_{\check{\Q}}\big[\check{Z}_t\i_{\{\tau_n\leq t\}}\big]=E_{\check{\Q}}\big[\check{Z}_{\tau_n}\i_{\{\tau_n\leq t\}}\big]\leq\frac{1}{n}.\]
We required the uniform integrability of $Z$ for the \emph{optional stopping theorem} in the penultimate step; see Theorem~II.3.2 in \cite{revuzyor} and the integral counterexample thereafter.\\
\emph{Proof of 2}. By construction, it holds
\begin{equation}\label{eq:proofinvconstr}E_\Q\big[H_t\big]=E_{\check{\Q}}\big[\check{Z}_tH_t\big]\end{equation}
for all $\Q$-integrable $\M{F}_t$-measurable random variables $H_t$. Let $A\in\M{F}_t$. If we set $H_t:=Z_t\i_A$, then we can write $\check{\Q}\big[A\cap\{\tau>t\}\big]=E_{\check{\Q}}\big[\check{Z}_tZ_t\i_{A\cap\{\tau>t\}}\big]=E_\Q\big[Z_t\i_{A\cap\{\tau>t\}}\big]=E_\Q\big[Z_t\i_A\big]$. In the first equation, we exploited that $\check{Z}_t>0$ holds $\check{\Q}$-a.s.\ on $\{\tau>t\}$. Then, we used Formula~\eqref{eq:proofinvconstr}. Eventually, we could omit the restriction to the event $\{\tau>t\}$, because it has full $\Q$-mass anyway.\\
\emph{Proof of 3}. According to the first step of the proof, $Z$ as the inverse of $\check{Z}$ is well-defined $\Q$-a.s. Furthermore, $Z$ is a $\Q$-supermartingale since it holds for all $0\leq u\leq t<\infty$ and for all $A\in\M{F}_u$
\[E_{\Q}\big[Z_t\i_A\big]=\check{\Q}\big[A\cap\{\tau>t\}\big]\leq\check{\Q}\big[A\cap\{\tau>u\}\big]=E_\Q\big[Z_u\i_A\big].\]
\emph{Proof of 4}. The argument is motivated by Example~4.1 in \cite{pr}. If Condition~\eqref{eq:notjump} is not satisfied, then $Z$ cannot form a $\Q$-local martingale. In fact, any localising sequence that preserves the expectation at the stopping time under $\Q$ must remain finite with positive $\Q$-probability. This certainly contravenes the local martingale property of $Z$; see Example~4.1 in \cite{pr} for the exact details. If Condition~\eqref{eq:notjump} is met, then $(\tau_n\wedge n)_{n\in\N}$ constitutes a localising sequence. Indeed, let $\rho$ be an arbitrary bounded stopping time. All we need to show is that $Z_\rho^{\tau_n\wedge n}=Z_{\rho\wedge\tau_n\wedge n}$ is $\Q$-integrable and that $E_\Q\big[Z_\rho^{\tau_n\wedge n}\big]=E_\Q\big[Z_0^{\tau_n\wedge n}\big]$ holds for all $n\in\N$; e.g., see Theorem~II.3.5 in \cite{revuzyor}. What we already know is the validity of the generalised Föllmer property
\begin{equation}\label{eq:föllmerpr}\check{\Q}\big[A\cap\{\tau>\rho\}\big]=E_\Q\big[Z_\rho\i_A\big]\end{equation}
for all $A\in\M{F}_\rho:=\sigma\Big(\set{A\in\M{F}}{A\cap\{\rho\leq t\}\in\M{F}_t\textrm{ for all }t\geq 0}\Big)$ and all finite stopping times $\rho$; see Proposition~2.3 together with Definition~2.1 in \cite{pr}. Alternatively, see the first part in the proof of Lemma~\ref{lem:naqcheckb} below. On the one hand, $Z$ is a $\Q$-supermartingale. Thus, the optional stopping theorem yields
\begin{equation}\label{eq:invföll1}E_\Q\big[Z_\rho^{\tau_n\wedge n}\big]\leq E_\Q\big[Z_0\big]=1.\end{equation}
On the other hand, Condition~\eqref{eq:notjump} is equivalent to saying that $\check{\Q}\big[(\tau_n\wedge n)<\tau\big]=1$ for all $n\in\N$. Consequently, \eqref{eq:föllmerpr} gives
\begin{equation}\label{eq:invföll2}E_\Q\big[Z_\rho^{\tau_n\wedge n}\big]=\check{\Q}\big[\tau>(\rho\wedge\tau_n\wedge n)\big]\geq\check{\Q}\big[\tau>(\tau_n\wedge\tau)\big]=1.\end{equation}
Combining \eqref{eq:invföll1} and \eqref{eq:invföll2} yields $E_\Q\big[Z_\rho^{\tau_n\wedge n}\big]\equiv 1$.\\
The last equivalence does not require separate attention. It follows straightforwardly with the same arguments as in the proof of Lemma~\ref{lem:charperfliq} below. This concludes our proof.
\end{bew}

\begin{remark}[Uniform Integrability of $\check{Z}$]\label{rmk:pathui}
\textup{
It needs to be noted that the uniform integrability of $\check{Z}$ in Lemma~\ref{lem:icsffm} is not a necessary condition. We only required it in the first step of the proof and in \eqref{eq:invföll1}. Generally, after having chosen a particular model, one may verify $\Q\big[\tau<\infty\big]=0$ and the local martingale property of $Z$ under $\Q$ alternatively; for instance, see Example~5.26 in \cite{krabi} for an illustration. In a non-pathological setting for which $\check{Z}$ is uniformly integrable, it generally holds $\check{\Q}\big[\tau<\infty\big]<1$. Indeed, if it held $\check{\Q}\big[\tau<\infty\big]=1$, then we would end up with the requisite $\check{Z}_t=E_{\check{\Q}}\big[\check{Z}_\infty\big|\M{F}_t\big]\equiv 0$ due to $\check{Z}_\infty=\lim_{t\to\infty}\check{Z}_t=0$; see Theorem~II.3.1 in \cite{revuzyor}.
}\hfill$\Box$
\end{remark}

\section{Stochastic Basis of the Enhanced FX-like Setting}\label{sec:basis}

Let $(\Omega,\M{F},\F)$ with $\F=(\M{F}_t)_{t\geq 0}$ denote a filtered space that is equipped with two exogenously given probability measures $\Q$ and $\check{\Q}$, where the local relation $\Q\big|_{\M{F}_t}\ll\check{\Q}\big|_{\M{F}_t}$ holds for every $t\geq 0$. The stochastic basis carries three $\F$-adapted families of zero-coupon bond price processes $\big\{P(t,T)\big\}_{0\leq t\leq T<\infty}$, $\big\{\tilde{P}(t,T)\big\}_{0\leq t\leq T<\infty}$ and $\big\{Q(t,T)\big\}_{0\leq t\leq T<\infty}$, such that the familiar payoff relations \eqref{def:payoffp} and \eqref{def:payoffq} from the FX-like approach are satisfied. The recovery rate process $S=(S_t)_{t\geq 0}$ is still defined via $S_t:=\tilde{P}(t,t)$. Furthermore, the stochastic basis carries the two genuine bank account num\'eraires $B=(B_t)_{t\geq 0}$ and $\check{B}=(\check{B})_{t\geq 0}$ from the domestic and the foreign market respectively. We restrict ourselves to a particular setting. The first two items are of a technical nature. The last one is motivated by the observations from Example~\ref{bsp:kst}.

\begin{ass}[The Enhanced FX-like Setting]\label{ass:enhFXsetting}
\ 
\begin{enumerate}
\item $\F$ satisfies the usual conditions under $\check{\Q}$.
\item All involved processes $\big(P(t,T)\big)_{0\leq t\leq T}$, $\big(\tilde{P}(t,T)\big)_{0\leq t\leq T}$, $\big(Q(t,T)\big)_{0\leq t\leq T}$ for any $T\geq 0$ as well as $S$, $B$ and $\check{B}$ are non-negative $\check{\Q}$-a.s. Moreover, they all admit $\check{\Q}$-indistinguishable c\`adl\`ag versions.
\item Concerning absence of arbitrage, the discounted price processes admit the following properties:
\begin{enumerate}
\item $\big({B_t}^{-1}P(t,T)\big)_{0\leq t\leq T}$ defines a $\Q$-local martingale for each maturity $T\geq 0$.
\item $\big({B_t}^{-1}\tilde{P}(t,T)\big)_{0\leq t\leq T}$ defines a $\Q$-local martingale for each maturity $T\geq 0$.
\item $Z=(Z_t)_{t\geq 0}$ with $Z_t:=\frac{S_t\check{B}_t}{S_0B_t}$ defines a $\Q$-local martingale and $\check{B}$ remains finite perpetually as seen under $\Q$. $\check{B}$ may nonetheless diverge and reach the cemetery state $\{+\infty\}$ in finite time as seen under $\check{\Q}$.
\item Let the \emph{explosion time} $\tau$ be defined as before in \eqref{eq:explosion}. Then, with an abuse of notation, the process $\check{Z}=(\check{Z}_t)_{t\geq 0}$ with $\check{Z}_t:=\frac{1}{Z_t}\i_{\{\tau>t\}}$,
in the sense of \eqref{eq:invZ} is a true $\check{\Q}$-martingale and coincide with the density process of $\Q\big|_{\M{F}_t}$ with respect to its locally dominating counterpart $\check{\Q}\big|_{\M{F}_t}$.
\end{enumerate}
\end{enumerate}
\end{ass}

The following four lemmata illustrate some implications of Assumption~\ref{ass:enhFXsetting}.

\begin{lem}[Föllmer Pair]\label{lem:cqisföllmer}
Let Assumption~\ref{ass:enhFXsetting} be met. Then, $(\check{\Q},\tau)$ is a Föllmer pair for $Z$ in the sense of Definition~\ref{def:föllmerpair}.
\end{lem}

\begin{bew}
Since $Z$ is a $\Q$-local martingale, it does not explode in finite time. Thus, we have $\Q\big[\tau<\infty\big]=0$. Moreover, by item 3.~d) of Assumption~\ref{ass:enhFXsetting}, we have for all $t\geq 0$ and for all $\Q$-integrable $\M{F}_t$-measurable functions $H_t$
\begin{equation}\label{eq:bewcqisföllm} E_\Q\big[H_t\big]=E_{\check{\Q}}\big[\check{Z}_tH_t\big].\end{equation}
Let $A\in\M{F}_t$. If we choose $H_t:=Z_t\i_A$, then we get $\check{\Q}\big[A\cap\{\tau>t\}\big]=E_{\check{\Q}}\big[\check{Z}_tH_t\big]=E_\Q\big[Z_t\i_A\big]$, where we used Formula~\eqref{eq:bewcqisföllm} in the second equation. This yields the assertion.
\end{bew}\\

Lemma~\ref{lem:cqisföllmer} will help us in the next section to extend the idea of Example~\ref{bsp:kst} in a measurable way to arbitrary time instances $t\geq 0$.

\begin{lem}[Perfect Liquidity]\label{lem:charperfliq}
Let Assumption~\ref{ass:enhFXsetting} be met. Then, the following statements are equivalent:
\begin{enumerate}
\item $\Q\big|_{\M{F}_t}\approx\check{\Q}\big|_{\M{F}_t}$ for each $t\geq 0$.
\item $\check{\Q}\big[\tau<\infty\big]=0$.
\item $Z$ is a true $\Q$-martingale.
\end{enumerate}
\end{lem}

\begin{bew}
"1. $\Longrightarrow$ 2." As shown in the proof of the previous lemma, $\{\tau\leq t\}$ is a $\Q$-nullset for all $t\geq 0$. Due to the assumed local equivalence, $\{\tau\leq t\}$ is also a $\check{\Q}$-nullset for all $t\geq 0$. Consequently, as $\{\tau\leq n\}_{n=1}^\infty$ is an increasing sequence and $\check{\Q}$ is $\sigma$-additive, we have
\[\check{\Q}\big[\tau<\infty\big]=\check{\Q}\bigg[\bigcup_{n=1}^\infty\{\tau\leq n\}\bigg]=\lim_{n\to\infty}\check{\Q}\big[\tau\leq n\big]=0.\]
"2. $\Longrightarrow$ 3." If $\{\tau<\infty\}$ is a $\check{\Q}$-nullset, then $Z\check{Z}$ becomes $\check{\Q}$-indistinguishable from a constant process at the level $1$. This obviously forms a martingale under $\check{\Q}$. According to the Bayes formula for conditional expectations, a process $X=(X_t)_{t\geq 0}$ is a $\Q$-martingale if and only if $\check{Z}X$ is a $\check{\Q}$-martingale; see Formula~\eqref{eq:bayespos}.\\
"3. $\Longrightarrow$ 1." We only need to show $\check{\Q}\big|_{\M{F}_t}\ll\Q\big|_{\M{F}_t}$. Let $A\in\M{F}_t$ be an arbitrary $\Q$-nullset. Firstly, if $Z$ is a true $\Q$-martingale, then we have $t\longmapsto E_\Q\big[Z_t\big]\equiv 1$. Secondly, as shown in the previous lemma, $\check{\Q}$ is a Föllmer measure of $Z$. Thus, by definition, we have $\check{\Q}\big[\tau\leq t\big]=1-\check{\Q}\big[\Omega\cap\{\tau>t\}\big]=1-E_\Q\big[Z_t\i_\Omega\big]=1-E_\Q\big[Z_t\big]=0$ for any $t\geq 0$.
Now we can proceed similarly as in the first step of the proof in order to verify that $\{\tau<\infty\}$ is also a $\check{\Q}$-nullset. Therefore, we may easily conclude with $\check{\Q}\big[A\big]=\check{\Q}\big[A\cap\{\tau>t\}\big]=E_\Q\big[Z_t\i_A\big]=0$.
\end{bew}\\

As indicated by Example~\ref{bsp:kst}, the strict local martingale property of $Z$ features aspects of illiquidity. Thus, Lemma~\ref{lem:charperfliq} provides equivalent characterisations of a market that is equipped with perfect asset liquidity. In this case, we could just as well consider the general FX-like setting of Assumption~\ref{ass:gen} instead.

\begin{lem}[Incomplete Filtration]\label{lem:Fnotcomplete}
Let Assumption~\ref{ass:enhFXsetting} be met and let $Z$ be a strict $\Q$-local martingale. Then, $\F$ cannot be complete under $\Q$.
\end{lem}

\begin{bew}
We proceed by contradiction. If $\F$ was $\Q$-complete, then we would have $\{\tau\leq T\}\in\M{F}_0$ for all $T\geq 0$. This is because $\{\tau\leq T\}$ is contained in the $\Q$-nullset $\{\tau<\infty\}$. On the contrary, it exists a $T>0$ such that $\check{\Q}\big[\tau\leq T\big]>0$. Otherwise, if no such $T$ existed, then $Z$ would be a true $\Q$-martingale according to the previous lemma. However, this would be a contradiction to the premise that $Z$ is a strict $\Q$-local martingale. Thus, under the assumption that $\F$ was $\Q$-complete, it would hold $\check{\Q}$-a.s.\ $1=\frac{1}{Z_0}\i_{\{\tau> 0\}}=\check{Z}_0=E_{\check{\Q}}\big[\check{Z}_T\big|\M{F}_0\big]$, and, as $\i_{\{\tau>T\}}=\i_{\{\tau>T\}}\i_{\{\tau>T\}}$ and $\i_{\{\tau>T\}}$ is $\M{F}_0$-measurable, $1=E_{\check{\Q}}\big[\check{Z}_T\big|\M{F}_0\big]\i_{\{\tau>T\}}$. Combining these two representations of $1$ yields $\check{\Q}\big[\tau>T\big]=1$, which is obviously a contradiction to our choice of $T$.
\end{bew}\\

The argument in the proof of Lemma~\ref{lem:Fnotcomplete} is very intuitive. If the explosion time $\tau$ of the bubble $Z$ is already known beforehand, then $\Q$ and $\check{\Q}$ essentially have to be equivalent. Lemma~\ref{lem:Fnotcomplete} highlights that modelling asset liquidity in the proposed way involves certain technical obstacles. We can no longer assume that $\F$ fulfils the usual conditions under $\Q$; see also Example~2.8 in \cite{pr}. Nonetheless, if $\F$ were not complete under $\check{\Q}$, it could be augmented to an $\big(\M{F},\check{\Q}\big)$-complete $\overline{\F}$ straightforwardly. $\tau$ would remain an $\overline{\F}$-stopping time and $Z$ an $\big(\overline{\F},\Q\big)$-local martingale according to Lemma~1 in \cite{larsson}. Additionally, Equation~\eqref{eq:föllmer1} would easily extend its area of validity to all $A\in\overline{\M{F}}_t$. Thus, the first item of Assumption~\ref{ass:enhFXsetting} does not pose any problems.

The last lemma in this section says under what circumstances the two-price economy in the foreign market does not involve arbitrage opportunities, at least up to the default time $\tau$. If it also holds $Q(T,T)=\i_{\{\tau>T\}}$ $\check{\Q}$-a.s., which is not far-fetched in the light of Lemma~\ref{lem:counterint} below, then absence of arbitrage can even be guaranteed for all times.

\begin{lem}[Absence of Arbitrage]\label{lem:naqcheckb}
Let Assumption~\ref{ass:enhFXsetting} be satisfied. Moreover, let $(\sigma_n)_{n\in\N}$ be a localising sequence for $\big({B_t}^{-1}\tilde{P}(t,T)\big)_{0\leq t\leq T}$ under $\Q$, which satisfies the monotonous convergence property $\lim_{n\to\infty}\sigma_n=T$ also under $\check{\Q}$. Then $\big(\sigma_n)_{n\in\N}$ is also a $\check{\Q}$-localising sequence for the $\check{\Q}$-local martingale $\big({\check{B}_t}^{-1}Q(t,T)\i_{\{\tau>t\}}\big)_{0\leq t\leq T}$.
\end{lem}

\begin{remark}[Regularity Assumption]\label{rmk:notrelax}
\textup{
The additional assumption about $(\sigma_n)_{n\in\N}$ is necessary and, unfortunately, cannot be relaxed. It is just uncertain how the characteristic properties of $(\sigma_n)_{n\in\N}$ carry over when changing to the dominating measure $\check{\Q}$. They still prevail on $\{\tau=\infty\}$, but not necessarily on the $\Q$-nullset $\{\tau<\infty\}$. As a matter of fact, one cannot modify $(\sigma_n)_{n\in\N}$ on $\{\tau<\infty\}$ to $(\tilde{\sigma}_n)_{n\in\N}$ and still comply with the requirements $\lim_{n\to\infty}\tilde{\sigma}_n=T$ $\check{\Q}$-a.s.\ and $\{\tilde{\sigma}_n\leq t\}\in\M{F}_t$ for all $t\geq 0$ and all $n\in\N$. Lemma~\ref{lem:naqcheckb} holds naturally when one equips the enhanced FX-like setting with HJM-dynamics; see Section~6.5 in \cite{krabi}.
}\hfill$\Box$
\end{remark}

\begin{bew}
Similarly as in the proof of Theorem~2.1 in \cite{cfr}, one extends Formula~\eqref{eq:bayespos} to
\begin{equation}\label{eq:genbayesstop} E_{\check{\Q}}\big[H_{\sigma\wedge T}\i_{\{\tau>\sigma\wedge T\}}\big|\M{F}_t\big]=\frac{1}{Z_{\sigma\wedge t}}\i_{\{\tau>\sigma\wedge t\}}E_\Q\big[Z_{\sigma\wedge T}H_{\sigma\wedge T}\big|\M{F}_t\big]\end{equation}
for all stopping times $\sigma$. Indeed, by construction of the Föllmer measure, it holds $d\check{\Q}\big|_{\M{F}_{\tau_n\leli}}=Z_{\tau_n}d\Q\big|_{\M{F}_{\tau_n\leli}}$, where $Z_{\tau_n}$ is well-defined since the stopped process $\big(Z_t^{\tau_n}\big)_{t\geq 0}$ forms a uniformly integrable martingale; see also Lemma~A.3 in \cite{cfr}. Let $A\in\M{F}_{\sigma\wedge T}$, where
\[\M{F}_{\sigma_\wedge T}:=\sigma\Big(\set{A\in\M{F}}{A\cap\{\sigma\wedge T\leq t\}\in\M{F}_t\textrm{ for all }t\geq 0}\Big).\]
Then, we can write
\begin{align*}
\check{\Q}\big[A\cap\{\tau>\sigma\wedge T\}\big]&=\lim_{n\to\infty}\check{\Q}\big[A\cap\{\tau_n>\sigma\wedge T\}\big]\\
&=\lim_{n\to\infty}E_\Q\big[Z_{\tau_n}\i_{A\cap\{\tau_n>\sigma\wedge T\}}\big]\\
&=\lim_{n\to\infty}E_\Q\big[Z_{\sigma\wedge T}\i_{A\cap\{\tau_n>\sigma\wedge T\}}\big]\\
&=E_\Q\big[Z_{\sigma\wedge T}\i_A\big],
\end{align*}
where we utilised $\Q\big[\tau=\infty\big]=1$ and dominated convergence in the last equation. Having Formula~\eqref{eq:föllmer1} generalised to capped stopping times, we can proceed exactly as in the proof of Lemma~\ref{lem:genbayes} in order to derive \eqref{eq:genbayesstop}. Let $(\sigma_n)_{n\in\N}$ be a localising sequence for $\big({B_t}^{-1}\tilde{P}(t,T)\big)_{0\leq t\leq T}$ under $\Q$. Thereby, $(\sigma_n)_{n\in\N}$ is increasing with $\lim_{n\to\infty}\sigma_n=T$ $\Q$-a.s.\ and it holds
\begin{equation}\label{eq:martproptp} E_\Q\bigg[\frac{\tilde{P}(\sigma_n\wedge t,T)}{B_{\sigma_n\wedge t}}\bigg|\M{F}_u\bigg]=\frac{\tilde{P}(\sigma_n\wedge u,T)}{B_{\sigma_n\wedge u}}\end{equation}
$\Q$-a.s.\ for all $0\leq u\leq t\leq T$ and all $n\in\N$. Consequently, combining \eqref{eq:genbayesstop} and \eqref{eq:martproptp} yields
\begin{align*}
& E_{\check{\Q}}\bigg[\frac{Q(\sigma_n\wedge t,T)}{B_{\sigma_n\wedge t}}\i_{\{\tau>\sigma_n\wedge t\}}\bigg|\M{F}_u\bigg]\\
&\ =\frac{B_{\sigma_n\wedge u}}{S_{\sigma_n\wedge u}\check{B}_{\sigma_n\wedge u}}\i_{\{\tau>\sigma_n\wedge u\}}E_\Q\bigg[\frac{S_{\sigma_n\wedge t}\check{B}_{\sigma_n\wedge t}}{B_{\sigma_n\wedge t}}\frac{Q(\sigma_n\wedge t,T)}{\check{B}_{\sigma_n\wedge t}}\bigg|\M{F}_u\bigg]\\
&\ =\frac{B_{\sigma_n\wedge u}}{S_{\sigma_n\wedge u}\check{B}_{\sigma_n\wedge u}}\i_{\{\tau>\sigma_n\wedge u\}}E_\Q\bigg[\frac{\tilde{P}(\sigma_n\wedge t,T)}{B_{\sigma_n\wedge t}}\bigg|\M{F}_u\bigg]\\
&\ =\frac{B_{\sigma_n\wedge u}}{S_{\sigma_n\wedge u}\check{B}_{\sigma_n\wedge u}}\i_{\{\tau>\sigma_n\wedge u\}}\frac{\tilde{P}(\sigma_n\wedge u,T)}{B_{\sigma_n\wedge u}}\\
&\ =\frac{Q(\sigma_n\wedge u,T)}{B_{\sigma_n\wedge u}}\i_{\{\tau>\sigma_n\wedge u\}}
\end{align*}
$\check{\Q}$-a.s.\ for all $0\leq u\leq t\leq T$ and all $n\in\N$.
\end{bew}

\begin{remark}[Weak Ineligibility]\label{rmk:weakinel}
\textup{
In order to end up in a non-trivial case of the enhanced FX-like setting as under Assumption~\ref{ass:enhFXsetting}, $Z$ does not necessarily have to be ineligible in the sense that $Z$ is a strict local martingale under any equivalent separating measure; e.g., see Definition~4.11 in \cite{krabi}. We only require that $Z$ is at least a strict local martingale under the exogenously chosen risk-neutral reference measure $\Q$. As a consequence, the local absolute continuity of $\Q$ with respect to $\check{\Q}$ will also be strict. However, there may well exist a locally equivalent measure $\widehat{\Q}\approx\Q$ under which $Z$ forms a true martingale; e.g., see Example~5.5 in \cite{schweizerbubble}. Therefore, we slightly weaken the notion of ineligibility.
}\hfill$\Box$
\end{remark}

\section{The Illiquidity Premium in the Enhanced FX-like Setting}\label{sec:illprem}

Motivated by Example~\ref{bsp:kst}, we make the following definition.

\begin{definition}[Illiquidity Deflator]
Let the enhanced FX-like setting as specified in Assumption~\ref{ass:enhFXsetting} be given. We define the \emph{illiquidity deflator} as the $\Q$-local martingale $Z=(Z_t)_{t\geq 0}$ with
\[Z_t:=\frac{S_t\check{B}_t}{S_0B_t}.\]
\end{definition}

The illiquidity deflator $Z$ is vulnerable to hyperinflation as $\check{B}$ tends to explode once the recovery rate has depreciated. As long as full recovery is given, $Z$ may be interpreted as the limiting value process of roll-over strategies in defaultable zero-coupon bonds with declining holding periods; see the exposition in \cite{kst2013}. $Z$ deflates illiquidity since the resulting pricing machinery is conducted as if there was perfect liquidity in the market.

Let the explosion time $\tau$ be defined as in Equation~\eqref{eq:explosion}. A priori, the random variable $\frac{\check{B}_t}{\check{B}_T}$ for $0\leq t\leq T<\infty$ is only well-defined on the event $\{\tau>T\}$. Nonetheless, we can extend its domain beyond $\tau$ almost arbitrarily. Once the foreign bank account process is about to reach the cemetery state $\{+\infty\}$, we suspend the previous regime and replace $\check{B}$ by a suitable $(0,\infty)$-valued $\check{B}^\circ$. More precisely, as we cannot override the value of $\check{B}_t$ in hindsight if $\tau$ happens to lie within the range $(t,T]$, we consider for $0\leq t\leq T<\infty$ the stochastic discount factors
\begin{equation}\frac{\check{B}_t}{\check{B}_T}\i_{\{\tau>T\}}+\frac{\check{B}_t^\circ}{\check{B}_T^\circ}\i_{\{\tau\leq T\}}.\label{ans:stochdisc}\end{equation}
Still, we will stick to the notation $\frac{\check{B}_t}{\check{B}_T}$. Under $\Q$, the changes are concentrated on a nullset and raise no issues. Under $\check{\Q}$, the density $\check{Z}_t=\frac{1}{Z_t}\i_{\{\tau>t\}}$ for $t\geq 0$ will not be affected by the modification either. For instance, we may want to set $\frac{\check{B}_t}{\check{B}_T}\equiv 1$ whenever $T\geq\tau$. Since this may be too restrictive in some applications, we simply proclaim a general integrability condition. As we shall see below, the values of $\frac{\check{B}_t}{\check{B}_T}$ beyond $\tau$ have a crucial impact on the term structure of illiquidity; see also the Examples~\ref{bsp:trivialext} and \ref{bsp:trivialext2} below. $\tau$ naturally describes the default time. If no counterparty is willing to lend capital under any circumstances, refinancing cost inevitably explode and the business model ceases to be viable. After this regime switch, $\check{B}^\circ$ accounts for the risky interest rate term structure of the post-bankruptcy era.

\begin{ass}[Integrability of the Stochastic Discount Factors]\label{ass:stochdiscinteg}
Let the setting of Assumption~\ref{ass:enhFXsetting} be given. In the sense of Ansatz~\eqref{ans:stochdisc}, we assume that the stochastic discount factor $\frac{\check{B}_t}{\check{B}_T}$ is $(0,\infty)$-valued $\check{\Q}$-a.s.\ and integrable with respect to $\check{\Q}$ for any $0\leq t\leq T<\infty$.
\end{ass}

\begin{definition}[Liquidity Adjusted Price and Illiquidity Premium]\label{def:illiqprem}
Let the setting of Assumption~\ref{ass:enhFXsetting} be given, Assumption~\ref{ass:stochdiscinteg} be fulfilled and $0\leq t\leq T<\infty$. We define the $t$-\emph{liquidity adjusted price} of a foreign zero-coupon bond maturing at time $T$ as
\[\check{Q}(t,T):=E_{\check{\Q}}\bigg[\frac{\check{B}_t}{\check{B}_T}\bigg|\M{F}_t\bigg].\]
The price difference $\mathbb{L}(t,T):=S_t\check{Q}(t,T)-\tilde{P}(t,T)=S_t\big(\check{Q}(t,T)-Q(t,T)\big)$ is referred to as the \emph{illiquidity premium} of $\tilde{P}(t,T)$.
\end{definition}

$S_t\check{Q}(t,T)$ may be interpreted as a domestic fair $t$-value of a defaultable zero-coupon bond seen from a foreign investor. Thus, its deviance from $\tilde{P}(t,T)$ is a natural candidate in order to quantify asset liquidity. A positive illiquidity premium $\mathbb{L}(t,T)$ infers an illiquid zero-coupon bond $\tilde{P}(t,T)$, whereas a negative illiquidity premium relates to hyperliquidity; see below for further clarification. It needs to be noted that the foreign market may only be free of arbitrage under $\check{\Q}$ within the stochastic interval $[0,\tau)$. $\mathbb{L}(t,T)$ is mainly of interest on $\{\tau>t\}$. After $\tau$, all obligations are unwinded and the hitherto existing liquidity framework becomes redundant.

\begin{remark}[Mathematical Concept of Illiquidity]\label{rmk:mathilliq}
\textup{
Limited institutional liquidity at time $t$ goes along with a low asset liquidity of $\tilde{P}(t,t+dt)$, that inevitably affects the recovery rate adversely. As the classical FX-like approach covers both default and migration risk, the enhanced FX-like setting unites the two aspects of asset and institutional liquidity. What is more, since the recovery rate $S$ enters the illiquidity deflator $Z$, the enhanced FX-like setting features an interdependence between credit and liquidity risk. Therefore, we deem the ineligibility of $Z$ the right mathematical concept to describe illiquidity. If required, aspects of liquidity can be analysed on a stand-alone basis simply by setting the recovery rate $S\equiv 1$; however these considerations are delicate due to inherent arbitrage. One rather interprets the impact of illiquidity as a deviance from the intrinsic economic value. Consequently, it suffices to consider one term structure together with an associated num\'eraire.
}\hfill$\Box$
\end{remark}

\begin{definition}[Illiquidity Factor]\label{def:illfact}
The term-dependent ratio
\[\check{\Xi}(t,T)=\frac{Q(t,T)}{\check{Q}(t,T)}\]
for $0\leq t\leq T<\infty$ is referred to as \emph{illiquidity factor}. $\check{\Xi}(t,T)<1$ features illiquidity, whereas $\check{\Xi}(t,T)=1$ describes an equilibrium between supply and demand.
\end{definition}

We tacitly assume that the term structure $\big\{\check{Q}(t,T)\big\}_{0\leq t\leq T<\infty}$ is strictly positive $\check{\Q}$-a.s. Thus, the illiquidity factors in Definition~\ref{def:illfact} are well-defined. This subtlety is not pedantic in the light of Lemma~\ref{lem:counterint} below. One may consider four different cases of the enhanced FX-like setting; see Table~\ref{tab:fcases}. Each case describes a distinct market situation.

\begin{table}[htp]
\centering
\begin{tabular}{|>{\centering}p{\widthof{model of the second kind}}||>{\centering}p{\widthof{model of the second kind}}|c|}
\hline&&\\
\multirow{2}{*}{}&$Z=(Z_t)_{t\geq 0}$ is a&$Z=(Z_t)_{t\geq 0}$ is a strict\\
&true $\Q$-martingale&$\Q$-local martingale\\
&&\\
\hhline{|=#=|=|}&&\\
\multirow{2}{*}{$\Big(\frac{\tilde{P}(t,T)}{B_t}\Big)_{0\leq t\leq T}$ is a true}&model of the $1^\textrm{st}$ kind,&model of the $2^\textrm{nd}$ kind,\\
&efficient market,&illiquid market,\\
$\Q$-martingale&$\mathbb{L}(t,T)\equiv0$&$\mathbb{L}(t,T)\geq0$\\
&&\\
\hline&&\\
\multirow{2}{*}{$\Big(\frac{\tilde{P}(t,T)}{B_t}\Big)_{0\leq t\leq T}$ is a strict}&model of the $3^\textrm{rd}$ kind,&model of the $4^\textrm{th}$ kind,\\
&hyperliquid market,&general market,\\
$\Q$-local martingale&$\mathbb{L}(t,T)\leq0$&$\mathbb{L}(t,T)$ state-dependent\\
&&\\
\hline
\end{tabular}
\vspace{1em}
\caption{This table provides an overview of the four distinct market situations in which the enhanced FX-like setting may be considered. In the models of the $1^\textrm{st}$ and the $3^\textrm{rd}$ kind, we have the local equivalence $\Q\big|_{\M{F}_t}\approx\check{\Q}\big|_{\M{F}_t}$. For those of the $2^\textrm{nd}$ and $4^\textrm{th}$ kind, it only holds $\Q\big|_{\M{F}_t}\ll\check{\Q}\big|_{\M{F}_t}$.}\label{tab:fcases}
\end{table}

\subsection{Model of the 1st Kind}

Let the setting of Assumption~\ref{ass:enhFXsetting} be given and let both $\big({B_t}^{-1}\tilde{P}(t,T)\big)_{0\leq t\leq T}$ and $Z=(Z_t)_{t\geq 0}$ be true $\Q$-martingales. In this case, the classical change of num\'eraire technique yields
\[\check{Q}(t,T)=E_{\check{\Q}}\bigg[\frac{\check{B}_t}{\check{B}_T}\bigg|\M{F}_t\bigg]=Q(t,T).\]
Hence, this model captures perfect liquidity with $\mathbb{L}(t,T)\equiv0$ and $\check{\Xi}(t,T)\equiv 1$. This is a special case of the general FX-like setting.

\subsection{Model of the 2nd Kind}\label{sec:2nd}

Likewise, let the setting of Assumption~\ref{ass:enhFXsetting} be given. Moreover, let $\big({B_t}^{-1}\tilde{P}(t,T)\big)_{0\leq t\leq T}$ be a true $\Q$-martingale, whereas $Z=(Z_t)_{t\geq 0}$ is a strict $\Q$-local martingale. If $\tau$ denotes the explosion time as defined in Equation~\eqref{eq:explosion}, then it holds by the Bayes formula~\eqref{eq:bayespos} both $\Q$-a.s.\ and $\check{\Q}$-a.s.
\begin{align}
S_t\check{Q}(t,T)&=S_tE_{\check{\Q}}\bigg[\frac{\check{B}_t}{\check{B}_T}\bigg|\M{F}_t\bigg]\nonumber\\
&\geq S_tE_{\check{\Q}}\bigg[\frac{\check{B}_t}{\check{B}_T}\i_{\{\tau>T\}}\bigg|\M{F}_t\bigg]\label{eq:argueordrel}\\
&=\i_{\{\tau>t\}}S_t\frac{B_t}{S_t\check{B}_t}E_{\Q}\bigg[\frac{S_T\check{B}_T}{B_T}\frac{\check{B}_t}{\check{B}_T}\bigg|\M{F}_t\bigg]=\i_{\{\tau>t\}}\tilde{P}(t,T).\nonumber
\end{align}
The liquidity adjusted prices also infer a non-defaultable term structure, since $\check{Q}(T,T)=1$ holds $\check{\Q}$-a.s.\ for all $T$. Due to $\Q\big[\tau=\infty\big]=1$, the illiquidity premia are non-negative $\Q$-a.s. This is, however, not necessarily the case $\check{\Q}$-a.s., since $\tilde{P}(t,T)$ still might exceed $S_t\check{Q}(t,T)$ on $\{\tau\leq t\}$.

\begin{remark}[Model of the $2^\textrm{nd}$ Kind]
\textup{
The presented concept of the illiquidity premium heavily relies on the premise that $\Q$ is a true martingale measure for the defaultable zero-coupon bonds. It cannot be relaxed without destroying the $\Q$-a.s.\ order relation $Q(t,T)\leq\check{Q}(t,T)$. If $\big({B_t}^{-1}\tilde{P}(t,T)\big)_{0\leq t\leq T}$ formed a strict $\Q$-local martingale and, hence, a $\Q$-supermartingale, then we would have both $\Q$-a.s.\ and $\check{\Q}$-a.s.
\[S_tE_{\check{\Q}}\bigg[\frac{\check{B}_t}{\check{B}_T}\i_{\{\tau>T\}}\bigg|\M{F}_t\bigg]=\frac{B_t}{\check{B}_t}\i_{\{\tau>t\}}E_{\Q}\bigg[\frac{S_T\check{B}_T}{B_T}\frac{\check{B}_t}{\check{B}_T}\bigg|\M{F}_t\bigg]\leq\i_{\{\tau>t\}}\tilde{P}(t,T),\]
where we used the Bayes formula~\eqref{eq:bayespos} in the first equation. Thus, the argument \eqref{eq:argueordrel} would not withstand any longer. Illiquidity as an asymmetry between supply and demand can attain two states. We deem the risk that you hardly find a buyer for certain corporate loans higher and more often present in the real world than the opposite situation in which an abundance of market participants is craving for a hardly available asset. Thus, we typically have $\mathbb{L}(t,T)\geq 0$, i.e., the corporate loans are traded below their intrinsic economic value. Equivalently, the issuer of the loans have to bear a higher interest rate burden such that market participants are willing to invest. All in all, the model of the $2^\textrm{nd}$ kind may be somehow considered as standard case in the presence of illiquidity. For instance, this can be utilised for the modelling of the interbank market; see Chapter~7 in \cite{krabi}.
}\hfill$\Box$
\end{remark}

\begin{example}[Flat Post-Default Curve]\label{bsp:trivialext}
\textup{
Let the enhanced FX-like setting of the $2^\textrm{nd}$ kind be given. If it holds $\frac{\check{B}_t}{\check{B}_T}\equiv 1$ for any $T\geq\tau$, then a straightforward calculation yields
\begin{align*}
\check{Q}(t,T)&=E_{\check{\Q}}\bigg[\frac{\check{B}_t}{\check{B}_T}\i_{\{\tau\leq T\}}\bigg|\M{F}_t\bigg]+E_{\check{\Q}}\bigg[\frac{\check{B}_t}{\check{B}_T}\i_{\{\tau>T\}}\bigg|\M{F}_t\bigg]\\
&=\check{\Q}\big[\tau\leq T\big|\M{F}_t\big]+Q(t,T)\i_{\{\tau>t\}}.
\end{align*}
Therefore, $\mathbb{L}(t,T)\i_{\{\tau>t\}}=S_t\check{\Q}\big[t<\tau\leq T\big|\M{F}_t\big]$, whereas the post-default illiquidity premium reduces to $\mathbb{L}(t,T)\i_{\{\tau\leq t\}}=\big(S_t-\tilde{P}(t,T)\big)\i_{\{\tau\leq t\}}$. The model features a liquidity premium that is, up to time $\tau$, proportional to the conditional default probability. After $\tau$, the liquidity premium is the deviance of the time value of the defaultable zero-coupon bond from the most current recovery rate. Thus, if $T\longmapsto\tilde{P}(t,T)$ does not become flat after $\tau$, bond holders must bear an illiquidity discount during the unwinding process. A pertinent model choice for $\check{Q}(t,T)$ may consist of an HJM-framework for $Q(t,T)$ and an intensity-based approach for $\tau$ under $\check{\Q}$; see Chapter~6 and Section~A.2 in \cite{krabi} for further details.}\hfill$\Box$
\end{example}

\begin{example}[Non-trivial Post-Default Curve]\label{bsp:trivialext2}
\textup{
Let the enhanced FX-like setting of the $2^\textrm{nd}$ kind be given. If the stochastic discount factors $\frac{\check{B}_t}{\check{B}_T}$ are replaced by $\frac{\check{B}_t^\circ}{\check{B}_T^\circ}$ for all $0\leq\tau\leq T<\infty$ and $t\leq T$, where $\check{B}^\circ$ itself induces the term structure $T\longmapsto\check{Q}^\circ(t,T)$ for $0\leq t\leq T<\infty$ and is conditionally independent from $\tau$, then similar calculations as in the previous example yield $\mathbb{L}(t,T)\i_{\{\tau>t\}}=S_t\check{Q}^\circ(t,T)\check{\Q}\big[t<\tau\leq T\big|\M{F}_t\big]$ and $\mathbb{L}(t,T)\i_{\{\tau\leq t\}}=S_t\big(\check{Q}^\circ(t,T)-Q(t,T)\big)\i_{\{\tau\leq t\}}$. In this case, the corresponding pre-default illiquidity premium is given by a product of the conditional default probability times the conversion of the term structure $T\longmapsto\check{Q}^\circ(t,T)$ into the domestic market. After $\tau$, the illiquidity premia become mere spreads.}\hfill$\Box$
\end{example}

\begin{remark}[Analytical Tractability of the Illiquidity Premium]\label{rmk:taunotcalculateable}
\textup{
In order to calculate an illiquidity premium as modelled in the previous two examples, one should be able to derive the cumulative distribution function of the explosion time $\tau$ under $\check{\Q}$. Exemplarily, the corresponding Laplace transform can be characterised if $Z$ is a one-dimensional diffusion; e.g., see \cite{karatzasruf} or \cite{mexico}. Other examples with semi-explicit formulae for the distribution of the explosion time can be constructed based on Section~6 of \cite{karatzasruf}. Unfortunately, only little is known in this regard if $Z$ follows general Itô-dynamics. This is in contrast to ordinary differential equations for which the explosion time is known explicitly; e.g., see \cite{groisman}. Facing that difficulty, the authors chose an indirect approach to model the illiquidity premium in Section~6.5 of \cite{krabi}.
}\hfill$\Box$
\end{remark}

\begin{prop}[Forward Measures]\label{prop:forfwdmeas2}
Let the setting of Assumption~\ref{ass:enhFXsetting} be given. Moreover, let $\big({B_t}^{-1}\tilde{P}(t,T)\big)_{0\leq t\leq T}$ be a true $\Q$-martingale. Then we have the absolute continuity $\tilde{\Q}^T\ll\check{\Q}^T$ and
\[\frac{d\tilde{\Q}^T}{d\check{\Q}^T}\bigg|_{\M{F}_t}=\frac{\check{Q}(0,T)}{Q(0,T)}\frac{Q(t,T)}{\check{Q}(t,T)}=\frac{\check{\Xi}(t,T)}{\check{\Xi}(0,T)}.\]
Particularly, $\big(\check{\Xi}(t,T)\big)_{0\leq t\leq T}$ is a $\check{\Q}^T$-martingale.
\end{prop}

\begin{bew}
It holds for any $0\leq t\leq T$
\[\frac{d\tilde{\Q}^T}{d\check{\Q}^T}\bigg|_{\M{F}_t}=\frac{d\tilde{\Q}^T}{d\Q}\bigg|_{\M{F}_t}\times\frac{d\Q}{d\check{\Q}}\bigg|_{\M{F}_t}\times\frac{d\check{\Q}}{d\check{\Q}^T}\bigg|_{\M{F}_t}=\frac{\tilde{P}(t,T)}{\tilde{P}(0,T)B_t}\times\frac{S_0B_t}{S_t\check{B}_t}\times\frac{\check{Q}(0,T)\check{B}_t}{\check{Q}(t,T)},\]
which yields the assertion.
\end{bew}\\

As it turns out, the term structure $T\longmapsto Q(t,T)$ is no longer default-free when it is considered under the dominating measure $\check{\Q}$. In fact, the corresponding payoffs $Q(T,T)=\i_{\{\tau>T\}}$ are of the all-or-nothing type. Full recovery is given throughout until but excluding $\tau$, and zero recovery prevails thereafter. Even though full consistency is ensured due to $\Q\big[\tau<\infty\big]=0$, this observation is counter-intuitive in the light of Paradigm~\ref{par:jt91}.

\begin{lem}[Illiquidity as an Invisible Default Event]\label{lem:counterint}
Let Assumption~\ref{ass:enhFXsetting} be satisfied. Moreover, let $\big({B_t}^{-1}\tilde{P}(t,T)\big)_{0\leq t\leq T}$ be a true $\Q$-martingale. Then it holds
\[Q(t,T)\i_{\{\tau\leq t\}}=\check{\Xi}(t,T)\i_{\{\tau\leq t\}}=0\]
$\check{\Q}$-a.s. for all $0\leq t\leq T<\infty$.
\end{lem}

\begin{remark}[Explicit Arbitrage]\label{rmk:explarbit}
\textup{
Let us assume that it holds $Q(t,T)<\check{Q}(t,T)$ and hypothetically (despite asset liquidity constraints) that both term structures are marketable. This opens the way for an explicit arbitrage under $\Q$, but not necessarily under $\check{\Q}$. Provided that the strategy is admissible, shorting $\check{Q}(t,T)$ and entering a long position in $Q(t,T)$ leaves you with a free lunch. As seen under $\Q$, the payoffs $\check{Q}(T,T)=Q(T,T)=1$ a.s.\ offset each other. Hence, the initial discrepancy can fully be consumed by the financial agent. In contrast, as seen under $\check{\Q}$, the smaller price $Q(t,T)$ accounts for the possible shortfall in $Q(T,T)=\i_{\{\tau>T\}}$ that may occur with strictly positive $\check{\Q}$-likelihood. Thus, the arbitrage opportunity presumably vanishes under $\check{\Q}$. Proposition~4.20 in \cite{krabi} tells us that exploiting this $\Q$-arbitrage is even optimal in some sense. It is certainly scalable arbitrarily; see also Example~\ref{bsp:explicit} below.
}\hfill$\Box$
\end{remark}

\begin{bew}
By construction of the enhanced FX-like approach, it holds under the stated premises for all $0\leq t\leq T<\infty$
\[0=\tilde{\Q}^T\big[\tau\leq t\big]=E_{\check{\Q}^T}\Bigg[\frac{d\tilde{\Q}^T}{d\check{\Q}^T}\bigg|_{\M{F}_t}\i_{\{\tau\leq t\}}\Bigg]=\check{\Xi}(0,T)^{-1}E_{\check{\Q}^T}\Big[\check{\Xi}(t,T)\i_{\{\tau\leq t\}}\Big],\]
where we used Proposition~\ref{prop:forfwdmeas2} in the last step. Therefore, it must hold $\check{\Q}^T$-a.s.\ $\check{\Xi}(t,T)\i_{\{\tau\leq t\}}=0$. Utilising the equivalence $\check{\Q}^T\approx\check{\Q}$ yields the assertion.
\end{bew}

\begin{lem}[Absence of Arbitrage]\label{lem:qstillmart}
Let the setting of Assumption~\ref{ass:enhFXsetting} be given. Moreover, let $\big({B_t}^{-1}\tilde{P}(t,T)\big)_{0\leq t\leq T}$ be a true $\Q$-martingale. In that case, the process $\big({\check{B}_t}^{-1}Q(t,T)\big)_{0\leq t\leq T}$ is a $\check{\Q}$-martingale.
\end{lem}

\begin{bew}
Using the previous lemma back and forth as well as the Bayes formula~\eqref{eq:bayespos}, we get $\check{\Q}$-a.s.\ for all $0\leq u\leq t\leq T<\infty$
\begin{align*}
E_{\check{\Q}}\bigg[\frac{Q(t,T)}{\check{B}_t}\bigg|\M{F}_u\bigg]=E_{\check{\Q}}\bigg[\frac{Q(t,T)}{\check{B}_t}\i_{\{\tau>t\}}\bigg|\M{F}_u\bigg]&=\frac{B_u}{S_u\check{B}_u}\i_{\{\tau>u\}}E_\Q\bigg[\frac{S_t\check{B}_t}{B_t}\frac{Q(t,T)}{\check{B}_t}\bigg|\M{F}_t\bigg]\\
&=\frac{B_u}{S_u\check{B}_u}\i_{\{\tau>u\}}\frac{\tilde{P}(u,T)}{B_u}=\frac{Q(u,T)}{\check{B}_u}.
\end{align*}
This proves the assertion.
\end{bew}

\begin{example}[Explicit Arbitrage]\label{bsp:explicit}
\textup{
The following construction is based on the article \cite{delbaen:bessel} that studies arbitrage opportunities in FX markets. The relevance of this article for our credit and liquidity risk setting is apparent.\\
Let $\check{W}=(\check{W})_{t\geq 0}$ be a Brownian motion on $(\Omega,\M{F},\F,\check{\Q})$ with its completed natural filtration $\F=(\M{F}_t)_{t\geq 0}$. The density process $\check{Z}=(\check{Z}_t)_{t\geq 0}$ with $\check{Z}_t:=1+\check{W}_{\tau\wedge t}$ and $\tau:=\inf\set{t>0}{\check{W}_t=-1}$ is a Brownian motion started at the level $1$ and stopped once it has hit the origin. We define the locally absolutely continuous measure $\Q\ll\check{\Q}$ on $\M{F}_\infty:=\bigvee_{t\geq 0}\M{F}_t$ via $\frac{d\Q}{d\check{\Q}}\big|_{\M{F}_t}:=\check{Z}_t$. Then it holds $\Q\big[\tau\leq T\big]=0$ for each $T\geq 0$ and
\[\tilde{W}_t:=\check{W}_t-\integ{0}{t}{\frac{1}{\check{Z}_u}}{u}\]
defines a $\Q$-Brownian motion. As seen under $\Q$, $\check{Z}$ is a \emph{Bessel process} of dimension three (\bes) started at $1$. Its inverse $Z_t={\check{Z}_t}^{-1}$ satisfies $dZ_t=-{Z_t}^2\,d\tilde{W}_t$ and is the stereotypical example of a strict $\Q$-local martingale. As exposed in \cite{delbaen:bessel}, the \bes\ process satisfies the no-arbitrage property with respect to simple integrands. However, it permits arbitrage with respect to general admissible integrands. Thus, while the submarket $(1,Z)$ with $Z_t=\frac{S_t\check{B}_t}{S_0B_t}$ and the num\'eraire $B=(B_t)_{t\geq 0}$ fulfils (NFLVR) under $\Q$, a riskless profit can be made after having conducted an undue change of num\'eraire to $Z$ resulting in $(1,\check{Z})$. This approach allows to specify the arbitrage strategy explicitly.\\
Let us fix a maturity $T>0$ for which $\check{\Q}\big[\tau\leq T\big]>0$ holds. The reflection principle for Brownian motion and the Markov property yield as derived in Section~5 of \cite{chautank}
\[\check{\Q}\big[\tau\leq T\big|\M{F}_t\big]=\begin{cases}1&\textrm{, on }\{\tau\leq t\},\\2\Phi\Big(-\frac{\check{Z}_t}{\sqrt{T-t}}\Big)&\textrm{, on }\{\tau> t\}.\end{cases}\]
Applying Itô's formula results on $\{\tau>t\}$ in the replication strategy
\[\check{\Q}\big[\tau\leq T\big|\M{F}_t\big]=\underbrace{\check{\Q}\big[\tau\leq T\big]}_{=2\Phi\big(-\frac{1}{\sqrt{T}}\big)<1}-\sqrt{\frac{2}{\pi}}\integ{0}{\tau\wedge t}{\frac{1}{\sqrt{T-u}}e^{-\frac{1}{2}\frac{{\check{Z}_u}^2}{T-u}}}{\check{Z}_u}.\]
$a_T:=\check{\Q}\big[\tau>T\big]/\check{\Q}\big[\tau\leq T\big]=1/2\Phi\big(-\frac{1}{\sqrt{T}}\big)-1$ is a strictly positive constant. The payoff $f:=\i_{\{\tau>T\}}-a_T\i_{\{\tau\leq T\}}$ can be perfectly replicated started from zero initial wealth. Indeed, the corresponding self-financing delta hedging strategy $H=(H_t)_{0\leq t\leq T}$ is given by
\[H_t=(1+a_T)\sqrt{\frac{2}{\pi}}\frac{1}{\sqrt{T-t}}e^{-\frac{1}{2}\frac{{\check{Z}_t}^2}{T-t}}.\]
The strategy certainly is $a_T$-admissible with respect to the num\'eraire $Z$ both under $\Q$ and $\check{\Q}$; otherwise, there would be an arbitrage opportunity. As seen under $\Q$, it results a.s.\ in the riskless payoff $\i_{\{\tau>T\}}=1$. In the traditional perspective with respect to the num\'eraire $B$, the strategy $H$ is not admissible. Indeed, unless $Z$ is a true $\Q$-martingale, which is prevented by the well-posedness of $a_T$, shorting $Z_t\check{\Q}\big[\tau\leq T\big|\M{F}_t\big]$ cannot be bounded from below. Indeed, the Bayes formula~\eqref{eq:bayespos} says $Z_t\check{\Q}\big[\tau\leq T\big|\M{F}_t\big]=Z_t-\i_{\{\tau>t\}}E_\Q\big[Z_T\big|\M{F}_t\big]$. The subtrahend is a.s.\ bounded from above by one, whereas $Z$ exceeds any level with positive $\Q$-probability.
}\hfill$\Box$
\end{example}

\begin{remark}[Exogeneity of the Foreign Num\'eraire]
\textup{
In fact, the foreign bank account process $\check{B}=(\check{B}_t)_{t\geq 0}$ in Example~\ref{bsp:explicit} is somehow exogenously given and not the infinitesimal roll-over of $\big\{Q(t,T)\big\}_{0\leq t\leq T<\infty}$. As such, it would often imply finite variation sample paths, where as the mentioned $\check{B}$ is generated beyond the short rate paradigm. If both $B$ and $\check{B}$ were inherited from short rate processes $r=(r_t)_{t\geq 0}$ and $\check{r}=(\check{r}_t)_{t\geq 0}$ respectively, then the resulting $\Q$-dynamics of $S=(S_t)_{t\geq 0}$ would be $dS_t=S_t(r_t-\check{r}_t)\,dt-S_tZ_t\,d\tilde{W}_t$. In order to keep the recovery rate $S$ within the target zone $(0,1]$, it must hold $B_tZ_t\leq\check{B}_t$ $\Q$-a.s.\ for all $t\geq 0$. Maintaining both this condition and full analytical tractability appears hardly achievable. For instance, one might postulate $\check{B}_t=X_tB_tZ_t$ for some auxiliary Itô-process $X=(X_t)_{t\geq 0}$ with state space $[1,\infty)$. The diffusion part of $X$ must be $X_tZ_t\,d\tilde{W}_t$ and needs to be compensated in the drift accordingly in order to ensure $X_t\geq 1$. Keeping this stochastic volatility process in the upper half-space is not trivial. In order to circumvent this perplexity, the authors propose an indirect HJM-approach in Section~6.5 of \cite{krabi}. To this end, the model features are inspired by the enhanced FX-like approach, but the characteristic of the foreign bank account is only secondary. In contrast, the next example presents a tractable case, where the foreign bank account coincides with the infinitesimal roll-over.
}\hfill$\Box$
\end{remark}

\begin{example}[Pure Illiquidity]
\textup{
This model is a modification of Example~8.1 in \cite{kst2013}. We consider the limiting case $S\equiv 1$, i.e., default risk is disabled. Thus, the term structures $\big\{P(t,T)\}_{0\leq t\leq T<\infty}$, $\big\{\tilde{P}(t,T)\}_{0\leq t\leq T<\infty}$ and $\big\{Q(t,T)\}_{0\leq t\leq T<\infty}$ all coincide. Let $x\in\R^4\setminus\{0\}$, $f:[0,\infty)\longrightarrow(0,\infty)$ denote a strictly positive, deterministic, c\`adl\`ag function and $\tilde{W}=(\tilde{W}_t)_{t\geq 0}$ be a four-dimensional Brownian motion on some filtered probability $(\Omega,\M{F},\F,\Q)$ with $\F=(\M{F}_t)_{t\geq 0}$ satisfying the usual conditions. We interpret the auxiliary process $X=(X_t)_{t\geq 0}$ with
\[X_t:=\frac{\big\|x+\tilde{W}_t\big\|^2}{f(t)}\]
as the evolution of the market's growth optimal portfolio. Its inverse is a strict $\Q$-local martingale. Utilising $X$ as natural num\'eraire, the marketable zero-coupon bond prices fulfil
\[P(t,T)=E_\Q\bigg[\frac{X_t}{X_T}\bigg|\M{F}_t\bigg]=\frac{f(T)}{f(t)}\bigg(1-e^{-\frac{\|x+\tilde{W}_t\|^{-2}}{2(T-t)}}\bigg);\]
see (8.2) in \cite{kst2013}. Provided that the mesh size of the discretisations can be controlled globally (see Example~8.1 in \cite{kst2013} for the exact details), the infinitesimal roll-over is $\Q$-a.s.\ given by $\check{B}=(\check{B}_t)_{t\geq 0}$ with $\check{B}_t=\frac{f(0)}{f(t)}$. If the discount factors remain deterministic as seen under a dominating Föllmer measure $\check{\Q}$ for the strict $\Q$-local martingale $Z=(Z_t)_{t\geq 0}$ with $Z_t=\frac{\check{B}_t}{X_t}$, then we easily get
\[\check{P}(t,T)=E_{\check{\Q}}\bigg[\frac{\check{B}_t}{\check{B}_T}\bigg|\M{F}_t\bigg]=\frac{f(T)}{f(t)}.\]
Thus, the fundamental values for all future times are already fixed at time $t=0$. The only driver for random fluctuations is the level of liquidity. The corresponding illiquidity premium $\mathbb{L}(t,T)=\check{P}(t,T)-P(t,T)$ is positive for all $0\leq t<T<\infty$ and vanishes at maturity.
}\hfill$\Box$
\end{example}

A general construction scheme and further examples can be found in Section~5.4 of \cite{krabi}.

\subsection{Model of the 3rd Kind}

Let the setting of Assumption~\ref{ass:enhFXsetting} be given. This time, let $\big({B_t}^{-1}\tilde{P}(t,T)\big)_{0\leq t\leq T}$ be a strict $\Q$-local martingale, whereas $Z=(Z_t)_{t\geq 0}$ is a true $\Q$ martingale. By Fatou's Lemma, the discounted prices of the defaultable zero-coupon bonds also form $\Q$-supermartingales. Consequently, we have
\[E_\Q\bigg[\frac{B_tS_T}{B_T}\bigg|\M{F}_t\bigg]\leq \tilde{P}(t,T).\]
Therefore, by the classical change of num\'eraire technique, we have
\[\check{Q}(t,T)=E_{\check{\Q}}\bigg[\frac{\check{B}_t}{\check{B}_T}\bigg|\M{F}_t\bigg]=\frac{1}{S_t}E_\Q\bigg[\frac{B_tS_T}{B_T}\bigg|\M{F}_t\bigg]\leq Q(t,T).\]
We observe that the defaultable zero-coupon bonds are overpriced and traded above their intrinsic economic value. For instance, this market inefficiency may be driven by emotions or prestige, or the simple fact that the intrinsic economic value is not readily observable. More commonly, this situation may occur in a regime with an abundance of foreign investors who bear the additional cost due to (more than) offsetting convenience effects of the currency conversion. Hereby, the domestic currency of the model acts as a safe haven. This is why even negative nominal interest rates could be enforced in Switzerland over the last couple of years.

\begin{example}[Hyperliquidity]
\textup{
We consider the enhanced FX-like setting from Assumption~\ref{ass:enhFXsetting} in the limiting case $S\equiv 1$ and $\check{B}=B$ $\Q$-a.s. Furthermore, we set $B_t:=X_t$ and ${B_t}^{-1}P(t,T):={X_t}^{-1}$ for a $\bes$-process $X=(X_t)_{t\geq 0}$ starting at $X_0=1$. In this case, we have $Z\equiv 1$, $\Q=\check{\Q}$ and
\[\check{P}(t,T)=E_{\check{\Q}}\bigg[\frac{\check{B}_t}{\check{B}_T}\bigg|\M{F}_t\bigg]=E_\Q\bigg[\frac{X_t}{X_T}\bigg|\M{F}_t\bigg]=1-2\Phi\bigg(-\frac{X_t}{\sqrt{T-t}}\bigg)<1=P(t,T)\]
for all $0\leq t<T<\infty$; see page~69 in \cite{honourplaten}
}\hfill$\Box$
\end{example}

\subsection{Model of the 4th Kind}

Let the setting of Assumption~\ref{ass:enhFXsetting} be given. Lastly, let both $\big({B_t}^{-1}\tilde{P}(t,T)\big)_{0\leq t\leq T}$ and $Z=(Z_t)_{t\geq 0}$ be a strict $\Q$-local martingales. On the one hand, we have similarly as in the model of the $2^\textrm{nd}$ kind
\[\check{Q}(t,T)\geq\i_{\{\tau>t\}}\frac{1}{S_t}E_\Q\bigg[\frac{B_tS_T}{B_T}\bigg|\M{F}_t\bigg].\]
Particularly, the statement still holds $\Q$-a.s.\ if we drop the restriction to the event $\{\tau>t\}$. On the other hand, exactly the same argument as in the model of the $3^\textrm{rd}$ kind carries over
\[\frac{1}{S_t}E_\Q\bigg[\frac{B_tS_T}{B_T}\bigg|\M{F}_t\bigg]\leq Q(t,T).\]
The model structure only maintains a common lower bound for $Q(t,T)$ and $\check{Q}(t,T)$ under $\Q$. Without any further model assumptions, nothing can be said about the direction of the illiquidity premium. A priori, we are in a general market situation in which the sign of $\mathbb{L}(t,T)$ is state-dependent. The $\M{F}_t$-event $\big\{\mathbb{L}(t,T)<0\big\}$ relates to an exorbitant demand for loans maturing at time $T$.

\begin{remark}[Analytical Intractability of the Illiquidity Premium]
\textup{
It is worth mentioning that neither Example~\ref{bsp:trivialext} nor Example~\ref{bsp:trivialext2} can be considered in this general market situation; more precisely, their premises cannot ever be satisfied. As $\mathbb{L}(t,T)$ may attain negative values on $\{\tau>t\}$, then $\check{\Q}\big[t<\tau\leq T\big|\M{F}_t\big]$ would have to become negative as well. Obviously, this would be absurd.
}\hfill$\Box$
\end{remark}

\section{Conclusion}

This article introduces a new mathematical concept of illiquidity that goes hand in hand with credit risk. Utilising the FX-analogy, the recovery rate stands for both institutional liquidity and that of the lending market. Asset liquidity constraints are nothing else than a hidden default; one sees two prices for a certain good, but one cannot exploit the price difference. At the explosion time, nobody is willing to hold the considered asset regardless of the promised yield being beyond any rational level. This is the occurrence of total illiquidity and coincides with the default time. In this sense, credit and liquidity risk can be modelled essentially in the same way.

\bibliographystyle{amsplain}

\end{document}